\documentclass[12pt]{article}
\usepackage{fullpage,times,graphicx,amssymb,psfrag,setspace}

\newcommand{\BEAS}{\begin{eqnarray*}}
\newcommand{\EEAS}{\end{eqnarray*}}
\newcommand{\BEA}{\begin{eqnarray}}
\newcommand{\EEA}{\end{eqnarray}}
\newcommand{\BEQ}{\begin{equation}}
\newcommand{\EEQ}{\end{equation}}
\newcommand{\BIT}{\begin{itemize}}
\newcommand{\EIT}{\end{itemize}}
\newcommand{\BNUM}{\begin{enumerate}}
\newcommand{\ENUM}{\end{enumerate}}

\newcommand{\BA}{\begin{array}}
\newcommand{\EA}{\end{array}}

\newcommand{\ie}{{\it i.e.}}

\newcommand{\ones}{\mathbf 1}

\newcommand{\reals}{{\mbox{\bf R}}}

\newcommand{\symm}{{\mbox{\bf S}}}  

\newcommand{\Rank}{\mathop{\bf Rank}}
\newcommand{\Card}{\mathop{\bf Card}}
\newcommand{\Tr}{\mathop{\bf Tr}}

\newcommand{\Expect}{\mathop{\bf E{}}}

\newcommand{\argmin}{\mathop{\rm argmin}}

\newcommand{\argmax}{\mathop{\rm argmax}}

\usepackage[agsmcite]{harvard}
\renewcommand{\cite}{\citeasnoun}

\begin{document}

\title{Identifying Small Mean Reverting Portfolios}
\author{By Alexandre d'Aspremont\footnote{
ORFE, Princeton University, Princeton, NJ 08544, USA. Email: \texttt{aspremon@princeton.edu}}}
\maketitle

\begin{abstract}
Given multivariate time series, we study the problem of forming portfolios with maximum mean reversion while constraining the number of assets in these portfolios. We show that it can be formulated as a sparse canonical correlation analysis and study various algorithms to solve the corresponding sparse generalized eigenvalue problems. After discussing penalized parameter estimation procedures, we study the sparsity versus predictability tradeoff and the impact of predictability in various markets.
\end{abstract}

\section{Introduction}
Mean reversion has received a significant amount of attention as a classic indicator of predictability in financial markets and is sometimes apparent, for example, in equity excess returns over long horizons. While mean reversion is easy to identify in univariate time series, isolating portfolios of assets exhibiting significant mean reversion is a much more complex problem. Classic solutions include cointegration or canonical correlation analysis, which will be discussed in what follows.

One of the key shortcomings of these methods though is that the mean reverting portfolios they identify are dense, i.e. they include every asset in the time series analyzed. For arbitrageurs, this means that exploiting the corresponding statistical arbitrage opportunities involves considerable \emph{transaction costs}. From an econometric point of view, this also impacts the \emph{interpretability} of the resulting portfolio and the significance of the structural relationships it highlights. Finally, optimally mean reverting portfolios often behave like noise and sometimes vary well inside bid-ask spreads, hence do not form meaningful statistical arbitrage opportunities. 

Here, we would like to argue that seeking \emph{sparse} portfolios instead, i.e. optimally mean reverting portfolios with a few assets, solves many of these issues at once: fewer assets means less transaction costs and more interpretable results. In practice, the tradeoff between mean reversion and sparsity is often very favorable. Furthermore, penalizing for sparsity also makes sparse portfolios vary in a wider price range, so the market inefficiencies they highlight are more significant.

Remark that all statements we will make here on mean reversion apply symmetrically to \emph{momentum}. Finding mean reverting portfolios using canonical correlation analysis means minimizing predictability, while searching for portfolios with strong momentum can also be done using canonical correlation analysis, by \emph{maximizing} predictability. The numerical procedures involved are identical.

Mean reversion has of course received a considerable amount of attention in the literature, most authors, such as \cite{Fama88}, \cite{Pote88} among many others, using it to model and test for predictability in excess returns. Cointegration techniques (see \cite{Engl87}, and \cite{Alex99} for a survey of applications in finance) are often used to extract mean reverting portfolios from multivariate time series. Early methods relied on a mix of regression and \cite{Dick79} stationarity tests or \cite{Joha88} type tests but it was subsequently discovered that an earlier canonical decomposition technique due to \cite{Box77} could be used to extract cointegrated vectors by solving a generalized eigenvalue problem (see \cite{Bewl94} for a more complete discussion). 

Several authors then focused on the optimal investment problem when excess returns are mean reverting, with \cite{Kim96} and \cite{Camp99} or \cite{Wach02} for example obtaining closed-form solutions in some particular cases. \cite{Liu04} also study the optimal investment problem in the presence of a ``textbook'' finite horizon arbitrage opportunity, modeled as a Brownian bridge, while \cite{Jure06} study this same problem when the arbitrage horizon is indeterminate. \cite{Gate06} studied the performance of pairs trading, using pairs of assets as classic examples of structurally mean-reverting portfolios. Finally, the LTCM meltdown in 1998 focused a lot of attention on the impact of leverage limits and liquidity, see \cite{Gros92} or \cite{Xion01} for a discussion.

Sparse estimation techniques in general and the $\ell_1$ penalization approach we use here in particular have also received a lot of attention in various forms: variable selection using the LASSO (see \cite{Tibs96}), sparse signal representation using basis pursuit by \cite{Chen01a}, compressed sensing (see \cite{Dono05} and \cite{Cand05}) or covariance selection (see \cite{Bane07}), to cite only a few examples. 
A recent stream of works on the asymptotic consistency of these procedures can be found in \cite{Mein07}, \cite{Cand07}, \cite{Bane07}, \cite{Yuan07} or \cite{Roth07} among others.

In this paper, we seek to adapt these results to the problem of estimating sparse (i.e. small) mean reverting portfolios. Suppose that $S_{ti}$ is the value at time $t$ of an asset $S_i$ with $i=1,\ldots,n$ and $t=1,\ldots,m$, we form portfolios $P_t$ of these assets with coefficients $x_i$, and assume they follow an Ornstein-Uhlenbeck process given by:
\BEQ\label{eq:ou-intro}
dP_t= \lambda(\bar P- P_t) dt + \sigma dZ_t \quad \mbox{with } P_t=\sum_{i=1}^n x_i S_{ti}
\EEQ
where $Z_t$ is a standard Brownian motion. Our objective here is to maximize the mean reversion coefficient $\lambda$ of $P_t$ by adjusting the portfolio weights $x_i$, under the constraints that $\|x\|=1$ and that the cardinality of $x$, \ie~the number of nonzero coefficients in $x$, remains below a given $k>0$. 

Our contribution here is twofold. First, we describe two algorithms for extracting sparse mean reverting portfolios from multivariate time series. One is based on a simple greedy search on the list of assets to include. The other uses semidefinite relaxation techniques to directly get good solutions. Both algorithms use predictability in the sense of \cite{Box77} as a proxy for mean reversion in (\ref{eq:ou-intro}). Second, we show that penalized regression and covariance selection techniques can be used as preprocessing steps to simultaneously stabilize parameter estimation and highlight key dependence relationships in the data. We then study the sparsity versus mean reversion tradeoff in several markets, and examine the impact of portfolio predictability on market efficiency using classic convergence trading strategies.

The paper is organized as follows. In Section~\ref{s:decomp}, we briefly recall the canonical decomposition technique derived in \cite{Box77}. In Section~\ref{s:algos}, we adapt these results and produce two algorithms to extract small mean reverting portfolios from multivariate data sets. In Section~\ref{s:cov-est}, we then show how penalized regression and covariance selection techniques can be used as preprocessing tools to both stabilize estimation and isolate key dependence relationships in the time series. Finally, we present some empirical results in Section~\ref{s:res} on U.S. swap rates and foreign exchange markets.

\section{Canonical decompositions}
\label{s:decomp} We briefly recall below the canonical decomposition technique derived in \cite{Box77}. Here, we work in a discrete setting and assume that the asset prices follow a stationary vector autoregressive process with:
\BEQ\label{eq:ar1}
S_t= S_{t-1}A + Z_t,
\EEQ
where $S_{t-1}$ is the lagged portfolio process, $A\in\reals^{n\times n}$ and $Z_t$ is a vector of i.i.d. Gaussian noise with zero mean and covariance $\Sigma \in \symm^n$, independent of $S_{t-1}$. Without loss of generality, we can assume that the assets $S_t$ have zero mean. The canonical analysis in \cite{Box77} starts as follows. For simplicity, let us first suppose that $n=1$ in equation (\ref{eq:ar1}), to get:
\[
\Expect[S_t^2]=\Expect[(S_{t-1}A)^2]+\Expect{[Z_t^2]},
\]
which can be rewritten as $\sigma^2_t=\sigma^2_{t-1}+\Sigma$. \cite{Box77} measure the \emph{predictability} of stationary series by:
\BEQ\label{eq:lambda-pred}
\nu=\frac{\sigma^2_{t-1}}{\sigma^2_t}.
\EEQ
The intuition behind this variance ratio is very simple: when it is small the variance of the noise dominates that of $S_{t-1}$ and $S_t$ is almost pure noise, when it is large however, $S_{t-1}$ dominates the noise and $S_t$ is almost perfectly predictable. Throughout the paper, we will use this measure of predictability as a proxy for the mean reversion parameter $\lambda$ in~(\ref{eq:ou-intro}). Consider now a portfolio $P_t=S_t x$ with weights $x\in\reals^{n}$, using~(\ref{eq:ar1}) we know that $S_tx=S_{t-1}Ax+Z_tx$, and we can measure its predicability as:
\[
\nu(x)=\frac{x^T A^T \Gamma A x}{x^T \Gamma x},
\]
where $\Gamma$ is the covariance matrix of $S_t$. Minimizing predictability is then equivalent to finding the minimum generalized eigenvalue $\lambda$ solving:
\BEQ\label{eq:geneig}
\det(\lambda \Gamma - A^T \Gamma A) =0.
\EEQ
Assuming that $\Gamma$ is positive definite, the portfolio with minimum predictability will be given by $x=\Gamma^{-1/2}z$, where $z$ is the eigenvector corresponding to the smallest eigenvalue of the matrix:
\BEQ\label{eq:pred-mat}
\Gamma^{-1/2} A^T \Gamma A \Gamma^{-1/2}.
\EEQ
We must now estimate the matrix $A$. Following \cite{Bewl94}, equation (\ref{eq:ar1}) can be written:
\[
S_t = \hat S_t + \hat Z_t,
\]
where $\hat S_t$ is the least squares estimate of $S_t$ with $\hat S_t=S_{t-1}\hat A$ and we get:
\BEQ\label{eq:est-A}
\hat A = \left(S_{t-1}^TS_{t-1}\right)^{-1} S_{t-1}^TS_{t}.
\EEQ
The \cite{Box77} procedure then solves for the optimal portfolio by inserting this estimate in the generalized eigenvalue problem above.

\begin{figure}[p]
\begin{center}
\includegraphics[width=\textwidth]{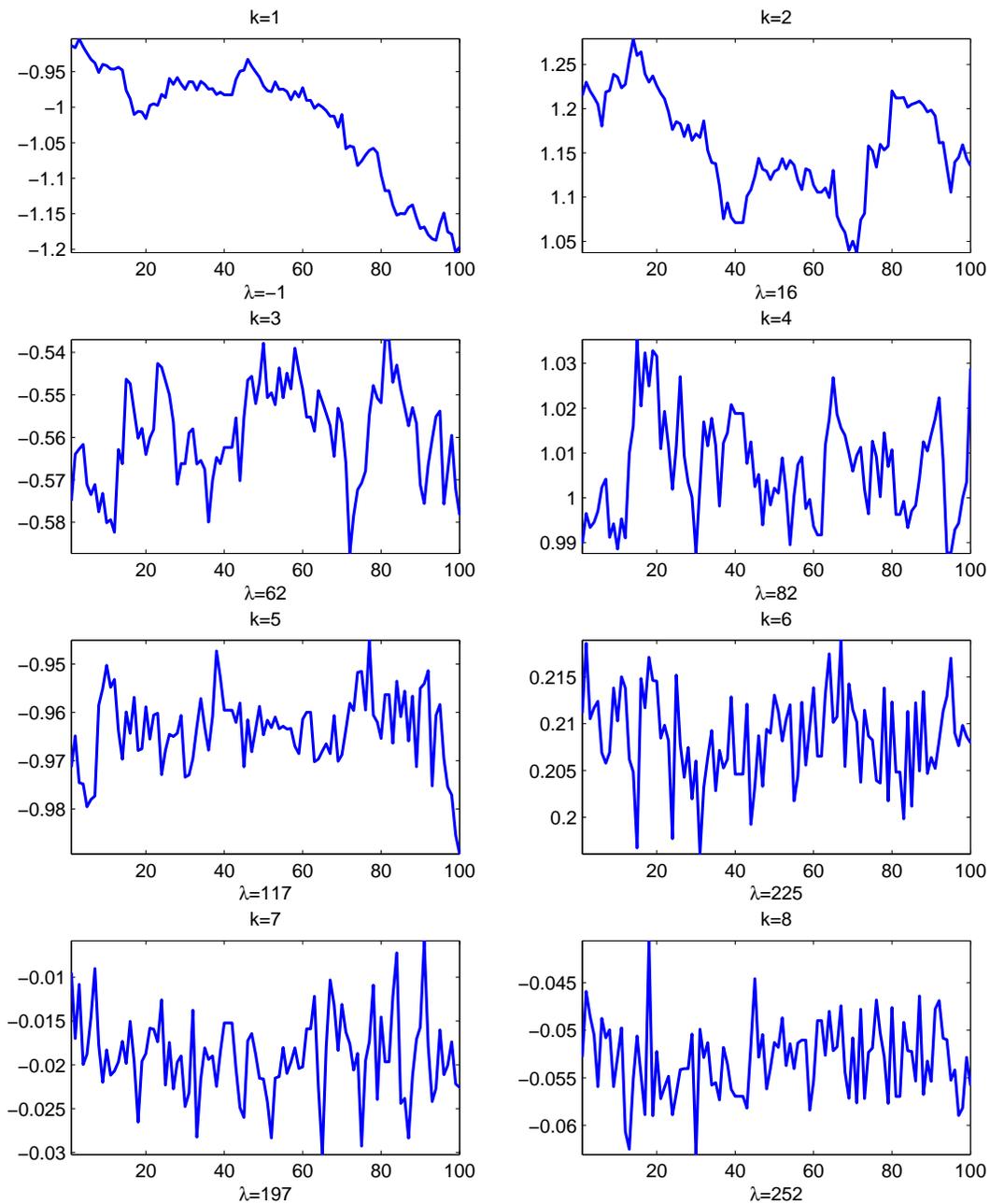} \caption{Box-Tiao decomposition on 100 days of U.S. swap rate data (in percent). The eight canonical portfolios of swap rates with maturities ranging from one to thirty years are ranked in decreasing order of predictability. The mean reversion coefficient $\lambda$ is listed below each plot. \label{fig:bt}}
\end{center}
\end{figure}

\paragraph{Box \& Tiao procedure.} Using the estimate (\ref{eq:est-A}) in (\ref{eq:pred-mat}) and the stationarity of $S_t$, the \cite{Box77} procedure finds linear combinations of the assets ranked in order of predictability by computing the eigenvectors of the matrix:
\BEQ\label{eq:bt}
\left(S_t^TS_t\right)^{-1/2}\left(\hat S_t^T \hat S_t \right)\left(S_t^TS_t\right)^{-1/2}
\EEQ
where $\hat S_t$ is the least squares estimate computed above. Figure \ref{fig:bt} gives an example of a \cite{Box77} decomposition on U.S. swap rates and shows eight portfolios of swap rates with maturities ranging from one to thirty years, ranked according to predictability. Table \ref{tab:swap-stats} shows mean reversion coefficient, volatility and the p-value associated with the mean reversion coefficient. We see that all mean reversion coefficients are significant at the 99\% level except for the last portfolio. For this highly mean reverting portfolio, a mean reversion coefficient of 238 implies a half-life of about one day, which explains the lack of significance on daily data.

\begin{table}[!h]
\small{
\begin{center}
\begin{tabular}{r|cccccccc}
Number of swaps: & 1&2&3&4&5&6&7&8\\
\hline
Mean reversion & 0.58 & 8.61 & 16.48 & 38.59 & 84.55 & 174.82 & 184.83 & 238.11 \\
P-value &0.00 & 0.00 & 0.00 & 0.00 & 0.00 & 0.00 & 0.00 & 0.51 \\
Volatility &0.21 & 0.28 & 0.34 & 0.14 & 0.10 & 0.09 & 0.07 & 0.07 \\  
\end{tabular}
\caption{Summary statistics for canonical U.S. swap portfolios: mean reversion coefficient, volatility and the p-value associated with the mean reversion coefficient for portfolio sizes ranging from one to eight.\label{tab:swap-stats}}
\end{center}}
\end{table}

\cite{Bewl94} show that the canonical decomposition above and the maximum likelihood decomposition in \cite{Joha88} can both be formulated in this manner. We very briefly recall their result below. 

\paragraph{Johansen procedure.} Following \cite{Bewl94}, the maximum likelihood procedure for estimating cointegrating vectors derived in \cite{Joha88} and \cite{Joha91} can also be written as a canonical decomposition \`a la \cite{Box77}. Here however, the canonical analysis is performed on the first order differences of the series $S_t$ and their lagged values $S_{t-1}$. We can rewrite equation~(\ref{eq:ar1}) as:
\[
\Delta S_t= Q S_{t-1} + Z_t
\]
where $Q=A-\mathbf{I}$. The basis of (potentially) cointegrating portfolios is then found by solving the following generalized eigenvalue problem:
\BEQ\label{eq:ml}
\lambda (S_{t-1}^TS_{t-1})-(S_{t-1}^T\Delta S_t (\Delta S_t^T\Delta S_t)^{-1}\Delta S_t^T S_{t-1})
\EEQ
in the variable $\lambda \in \reals$. 

\section{Sparse decomposition algorithms}
\label{s:algos}
In the previous section, we have seen that canonical decompositions can be written as generalized eigenvalue problems of the form:
\BEQ\label{eq:geneig}
\det(\lambda B-A)=0
\EEQ
in the variable $\lambda\in\reals$, where $A,B\in\symm^n$ are symmetric matrices of dimension $n$. Full generalized eigenvalue decomposition problems are usually solved using a QZ decomposition. Here however, we are only interested in extremal generalized eigenvalues, which can be written in variational form as:
\[
\lambda^{\mathrm{max}}(A,B)=\max_{x\in \scriptsize{\reals^n}} \frac{x^TAx}{x^TBx}.
\]
In this section, we will seek to maximize this ratio while constraining the cardinality of the (portfolio) coefficient vector $x$ and solve instead:
\BEQ \label{eq:sparse-eig}
\BA{ll}
\mbox{maximize} & {x^TAx}/{x^TBx} \\
\mbox{subject to} & \Card(x)\leq k\\
& \|x\|=1,
\EA\EEQ
where $k>0$ is a given constant and $\Card(x)$ is the number of nonzero coefficients in $x$. This will compute a sparse portfolio with maximum predictability (or momentum), a similar problem can be formed to minimize it (and obtain a sparse portfolio with maximum mean reversion). This is a hard combinatorial problem, in fact, \cite{Nata95} shows that sparse generalized eigenvalue problems are equivalent to subset selection, which is NP-hard. We can't expect to get optimal solutions and we discuss below two efficient techniques to get good approximate solutions.

\subsection{Greedy search}
\label{ss:greedy} Let us call $I_k$ the support of the solution vector $x$ given $k>0$ in problem (\ref{eq:sparse-eig}):
\[
I_k=\{i\in[1,n]:~x_i \neq 0\},
\]
by construction $|I_k|\leq k$. We can build approximate solutions to (\ref{eq:sparse-eig}) recursively in $k$. When $k=1$, we simply find $I_1$ as:
\[
I_1=\argmax_{i\in[1,n]} A_{ii}/B_{ii}.
\]
Suppose now that we have a good approximate solution with support set $I_k$ given by:
\[
x_{k}=\argmax_{\{x\in \scriptsize{\reals^n}:~x_{I_k^c}=0\}} \frac{x^TAx}{x^TBx},
\]
where $I^c_k$ is the complement of the set $I_k$. This can be solved as a generalized eigenvalue problem of size $k$. We seek to add one variable with index $i_{k+1}$ to the set $I_k$ to produce the largest increase in predictability by scanning each of the remaining indices in $I^c_k$. The index $i_{k+1}$ is then given by:
\[
i_{k+1}=\argmax_{i\in I^c_k} ~\max_{\{x\in \scriptsize{\reals^n}:~x_{J_i}=0\}}~ \frac{x^TAx}{x^TBx},\quad \mbox{where } J_i=I^c_{k}~\backslash~ \{i\},
\]
which amounts to solving $(n-k)$ generalized eigenvalue problems of size $k+1$. We then define:
\[
I_{k+1}=I_k \cup \{i_{k+1}\},
\] 
and repeat the procedure until $k=n$. Naturally, the optimal solutions of problem (\ref{eq:sparse-eig}) might not have increasing support sets $I_k \subset I_{k+1}$, hence the solutions found by this recursive algorithm are potentially far from optimal. However, the cost of this method is relatively low: with each iteration costing $O(k^2(n-k))$, the complexity of computing solutions for all target cardinalities $k$ is $O(n^4)$. This recursive procedure can also be repeated forward and backward to improve the quality of the solution.

\subsection{Semidefinite relaxation}
An alternative to greedy search which has proved very efficient on sparse maximum eigenvalue problems is to derive a convex relaxation of problem (\ref{eq:sparse-eig}). In this section, we extend the techniques of \cite{dasp04a} to formulate a semidefinite relaxation for sparse generalized eigenvalue problems in (\ref{eq:sparse-eig}):
\[\BA{ll}
\mbox{maximize} & {x^TAx}/{x^TBx} \\
\mbox{subject to} & \Card(x)\leq k\\
& \|x\|=1,
\EA\]
with variable $x\in\reals^n$. As in \cite{dasp04a}, we can form an equivalent program in terms of the matrix $X=xx^T\in\symm_n$:
\[\BA{ll}
\mbox{maximize} & {\Tr(AX)}/{\Tr(BX)} \\
\mbox{subject to} & \Card(X)\leq k^2\\
& \Tr(X)=1\\
& X\succeq 0,~\Rank(X)=1,
\EA\]
in the variable $X\in\symm_n$. This program is equivalent to the first one: indeed, if $X$ is a solution to the above problem, then $X \succeq 0$ and $\Rank(X)=1$ mean that we must have $X=xx^T$, while $\Tr(X)=1$ implies that $\|x\|=1$. Finally, if $X=xx^T$ then $\Card(X) \leq k^2$ is equivalent to $\Card(x) \leq k$. 

Now, because for any vector $u\in\reals^n$, $\Card(u)=q$ implies $\|u\|_1 \leq \sqrt{q} \|u\|_2$, we can replace the nonconvex constraint $\Card(X) \leq k^2$ by a weaker but convex constraint $\ones ^T |X| \ones \leq k$, using the fact that $\|X\|_F= \sqrt{x^Tx}=1$ when $X=xx^T$ and $\Tr(X)=1$. We then drop the rank constraint to get the following relaxation of (\ref{eq:sparse-eig}):
\BEQ\BA{ll}
\label{eq:max-meanrev-relax}
\mbox{maximize} & {\Tr(AX)}/{\Tr(BX)} \\
\mbox{subject to} & \ones^T|X|\ones \leq k\\
& \Tr(X)=1\\
& X\succeq 0,
\EA\EEQ
which is a quasi-convex program in the variable $X\in\symm_n$. After the following change of variables:
\[
Y=\frac{X}{\Tr(BX)}, \quad z=\frac{1}{\Tr(BX)},
\]
and rewrite (\ref{eq:max-meanrev-relax}) as:
\BEQ\BA{ll}
\label{eq:max-meanrev-change}
\mbox{maximize} & \Tr(AY) \\
\mbox{subject to} & \ones^T|Y|\ones - k z \leq 0\\
& \Tr(Y)-z=0\\
& \Tr(BY)=1\\
& Y\succeq 0,
\EA\EEQ
which is a semidefinite program (SDP) in the variables $Y\in\symm_n$ and $z\in\reals_+$ and can be solved using standard SDP solvers such as SEDUMI by \cite{Stur99} and SDPT3 by \cite{Toh96}. The optimal value of problem (\ref{eq:max-meanrev-change}) will be an upper bound on the optimal value of the original problem (\ref{eq:sparse-eig}). If the solution matrix $Y$ has rank one, then the relaxation is \emph{tight} and both optimal values are equal. When $\Rank(Y)>1$ at the optimum in (\ref{eq:max-meanrev-change}), we get an approximate solution to (\ref{eq:sparse-eig}) using the rescaled leading eigenvector of the optimal solution matrix $Y$ in (\ref{eq:max-meanrev-change}). The computational complexity of this relaxation is significantly higher than that of the greedy search algorithm in $\S \ref{ss:greedy}$. On the other hand, because it is not restricted to increasing sequences of sparse portfolios, the performance of the solutions produced is often higher too. Furthermore, the dual objective value produces an upper bound on suboptimality. Numerical comparisons of both techniques are detailed in Section \ref{s:res}.


\section{Parameter estimation}
\label{s:cov-est} The canonical decomposition procedures detailed in Section \ref{s:decomp} all rely on simple estimates of both the covariance matrix $\Gamma$ in (\ref{eq:pred-mat}) and the parameter matrix $A$ in the vector autoregressive model (\ref{eq:est-A}). Of course, both estimates suffer from well-known stability issues and a classic remedy is to penalize the covariance estimation using, for example, a multiple of the norm of $\Gamma$. In this section, we would like  to argue that using an $\ell_1$ penalty term to stabilize the estimation, in a procedure known as covariance selection, simultaneously stabilizes the estimate and helps isolate key idiosyncratic dependencies in the data. In  particular, covariance selection clusters the input data in several smaller groups of highly dependent variables among which we can then search for mean reverting (or momentum) portfolios. Covariance selection can then be viewed as a preprocessing step for the sparse canonical decomposition techniques detailed in Section \ref{s:algos}. Similarly, penalized regression techniques such as the LASSO by \cite{Tibs96} can be used to produce stable, structured estimates of the matrix parameter $A$ in the VAR model (\ref{eq:ar1}). 

\subsection{Covariance selection}
Here, we first seek to estimate the covariance matrix $\Gamma$ by maximum likelihood. Following \cite{Demp72}, we penalize the maximum-likelihood estimation  to set a certain number of coefficients in the inverse covariance matrix to zero, in a procedure known as \emph{covariance selection}. Zeroes in the inverse covariance matrix correspond to conditionally independent variables in the model and this approach can be used to simultaneously obtain a robust estimate of the covariance matrix while, perhaps more importantly, discovering \emph{structure} in the underlying graphical model (see \cite{Laur96} for a complete treatment). This tradeoff between log-likelihood of the solution and number of zeroes in its inverse (i.e. model structure) can be formalized in the following problem:
\BEQ \label{eq:sparseml-primal}
\max_X ~ \log \det X - \Tr(\Sigma X) - \rho \Card(X)\\
\EEQ
in the variable $X\in \symm_n$, where $\Sigma \in \symm_n$ is the sample covariance matrix, $\Card(X)$ is the cardinality of $X$, i.e. the number of nonzero coefficients in $X$ and $\rho >0$ is a parameter controlling the trade-off between likelihood and structure.

Solving the penalized maximum likelihood estimation problem in (\ref{eq:sparseml-primal}) both improves the stability of this estimation procedure by implicitly reducing the number of parameters and directly highlights structure in the underlying model. Unfortunately, the cardinality penalty makes this problem very hard to solve numerically. One solution developed in \cite{dAsp06b}, \cite{Bane07} or \cite{Frie07} is to relax the $\Card(X)$ penalty and replace it by the (convex) $\ell_1$ norm of the coefficients of $X$ to solve:
\BEQ \label{eq:sparseml-relax}
\max_X ~ \log \det X - \Tr(\Sigma X) - \rho \sum_{i,j=1}^n |X_{ij}|\\
\EEQ
in the variable $X \in \symm^n$. The penalty term involving the sum of absolute values of the entries of $X$ acts as a proxy for the cardinality: the function $\sum_{i,j=1}^n |X_{ij}|$ can be seen as the largest convex lower bound on $\Card(X)$ on the hypercube, an argument used by \cite{Boyd00} for rank minimization. It is also often used in regression and variable selection procedures, such as the LASSO by \cite{Tibs96}. Other permutation invariant estimators have been detailed in \cite{Roth07} for example.

\begin{figure}[tp]
\begin{center}
\psfrag{card}[t][b]{Cardinality}
\psfrag{sharpe}[b][t]{Sharpe Ratio}
\includegraphics[width=.42\textwidth]{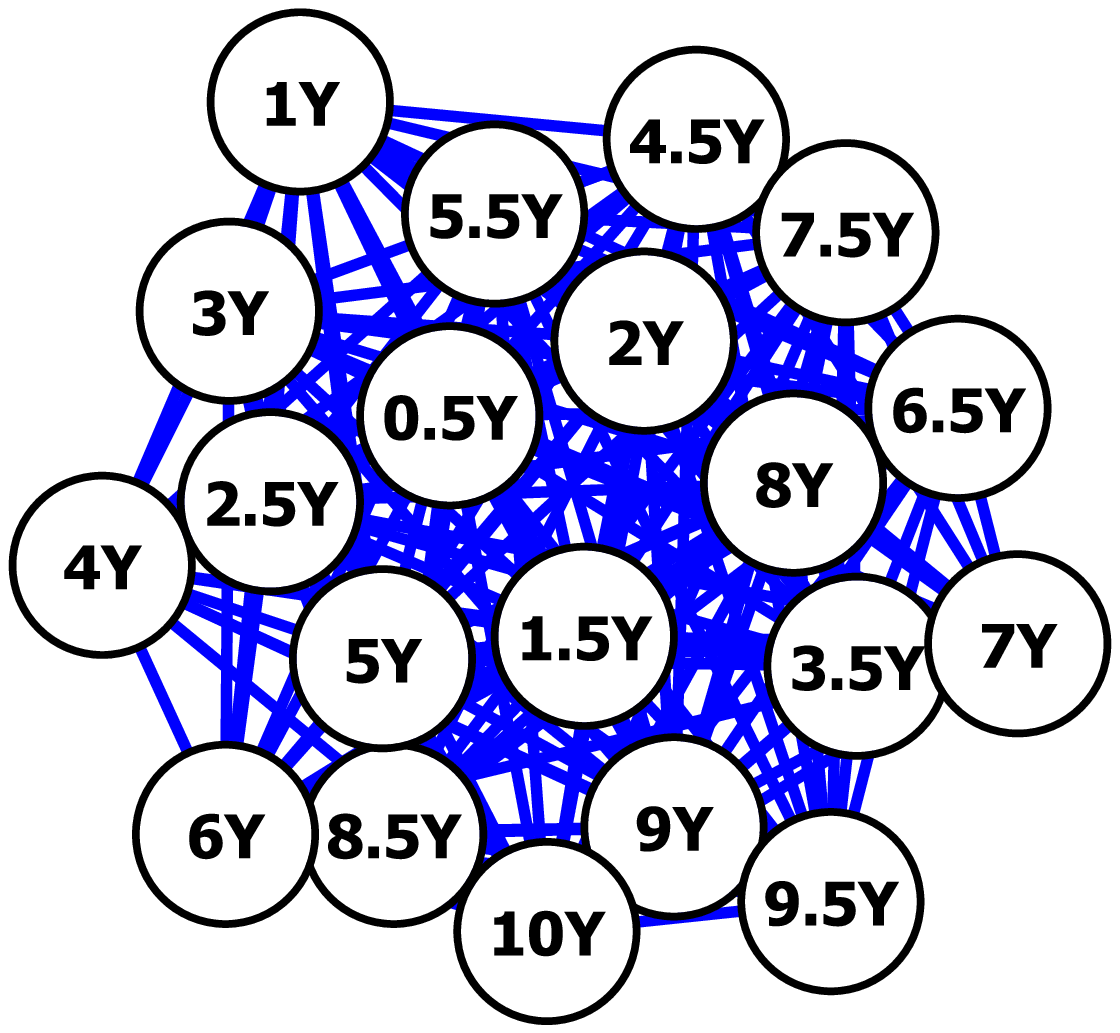} \quad
\includegraphics[width=.30\textwidth]{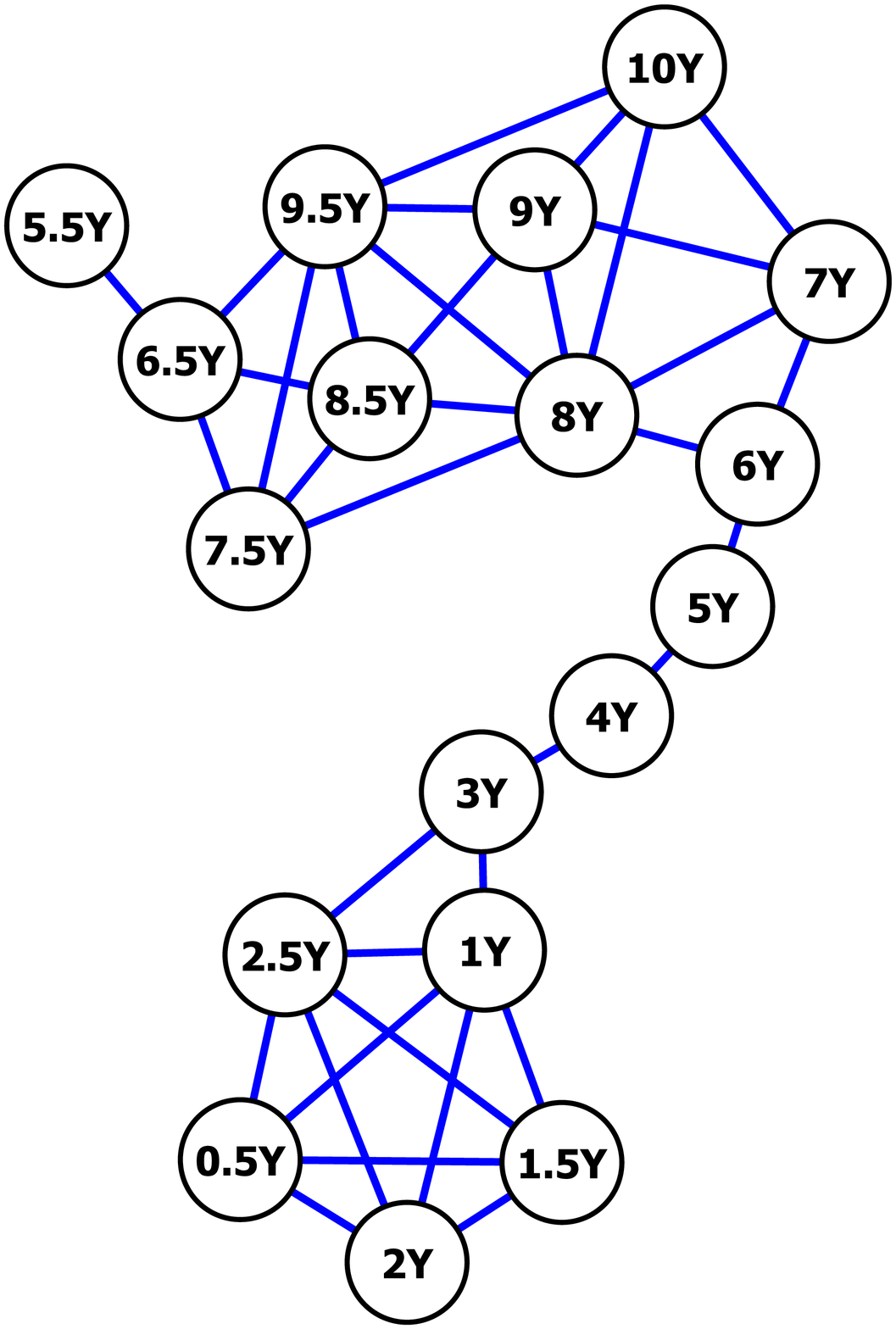}  \quad \quad\hfill 
\caption{\emph{Left:} conditional dependence network inferred from the pattern of zeros in the inverse swap covariance matrix. \emph{Right:} same plot, using this time the penalized covariance estimate with penalty $\rho=.1$ in the maximum likelihood estimation (\ref{eq:sparseml-relax}). \label{fig:swapnet}}
\end{center}
\end{figure}

In a Gaussian model, zeroes in the inverse covariance matrix point to variables that are conditionally independent, conditioned on all the remaining variables. This has a clear financial interpretation: the inverse covariance matrix reflects independence relationships between the \emph{idiosyncratic} components of asset price dynamics. In Figure \ref{fig:swapnet}, we plot the resulting network of dependence, or graphical model for U.S. swap rates. In this graph, variables (nodes) are joined by a link if and only if they are conditionally dependent. We plot the graphical model inferred from the pattern of zeros in the inverse sample swap covariance matrix (left) and the same graph, using this time the penalized covariance estimate in (\ref{eq:sparseml-relax}) with penalty parameter $\rho=.1$ (right). The graph layout was done using Cytoscape. Notice that in the penalized estimate, rates are clustered by maturity and the graph clearly reveals that swap rates are moving as a curve.

\subsection{Estimating structured VAR models}
In this section, using similar techniques, we show how to recover a sparse vector autoregressive model from multivariate data.

\paragraph{Endogenous dependence models.}
Here, we assume that the conditional dependence structure of the assets $S_t$ is purely \emph{endogenous}, i.e. that the noise terms in the vector autoregressive model~(\ref{eq:ar1}) are i.i.d. with:
\[
S_t= S_{t-1}A + Z_t,
\]
where $Z_t\sim\mathcal{N}(0,\sigma\mathbf{I})$ for some $\sigma>0$. In this case, we must have:
\[
\Gamma=A^T\Gamma A+\sigma\mathbf{I}
\]
since $A^T\otimes A$ has no unit eigenvalue (by stationarity), this means that:
\[
\Gamma/\sigma=(\mathbf{I}-A^T\otimes A^T)^{-1}\mathbf{I}
\]
where $A\otimes B$ is the Kronecker product of $A$ and $B$, which implies:
\[
A^TA=\mathbf{I}-\sigma \Gamma^{-1}.
\]
We can always choose $\sigma$ small enough so that $\mathbf{I}-\sigma \Gamma^{-1}\succeq 0$. This means that we can directly get $A$ as a matrix square root of $(\mathbf{I}-\sigma \Gamma^{-1})$. Furthermore, if we pick $A$ to be the Cholesky decomposition of $(\mathbf{I}-\sigma \Gamma^{-1})$, and if the graph of $\Gamma$ is chordal (i.e. has no cycles of length greater than three) then there is a permutation of the variables $P$ such that the Cholesky decomposition of $P\Gamma P^T$, and the upper triangle of $P\Gamma P^T$ have the same pattern of zeroes (see \cite{Werm80} for example). In Figure \ref{fig:chordal}, we plot two dependence networks, one chordal (on the left), one not (on the right). In this case, the structure (pattern of zeroes) of $A$ in the VAR model (\ref{eq:est-A}) can be directly inferred from that of the penalized covariance estimate.

\cite[\S2.4]{Gilb94} also shows that if $A$ satisfies $A^TA=\mathbf{I}-\sigma \Gamma^{-1}$ then, barring numerical cancellations in $A^TA$, the graph of $\Gamma^{-1}$ is the intersection graph of $A$ so:
\[
(\Gamma^{-1})_{ij}=0 ~\Longrightarrow~ A_{ki}A_{kj}=0,~\mbox{for all } k=1,\ldots,n.
\]
This means in particular that when the graph of $\Gamma$ is disconnected, then the graph of $A$ must also be disconnected along the same clusters of variables, i.e. $A$ and $\Gamma$ have identical block-diagonal structure. In $\S \ref{ss:pen-decomp}$, we will use this fact to show that when the graph of $\Gamma$ is disconnected, optimally mean reverting portfolios must be formed exclusively of assets within a single cluster of this graph.

\begin{figure}[tp]
\begin{center}
\psfrag{card}[t][b]{Cardinality}
\psfrag{sharpe}[b][t]{Sharpe Ratio}
\includegraphics[width=.28\textwidth]{./figures/RatesNetNew.eps} \quad
\includegraphics[width=.43\textwidth]{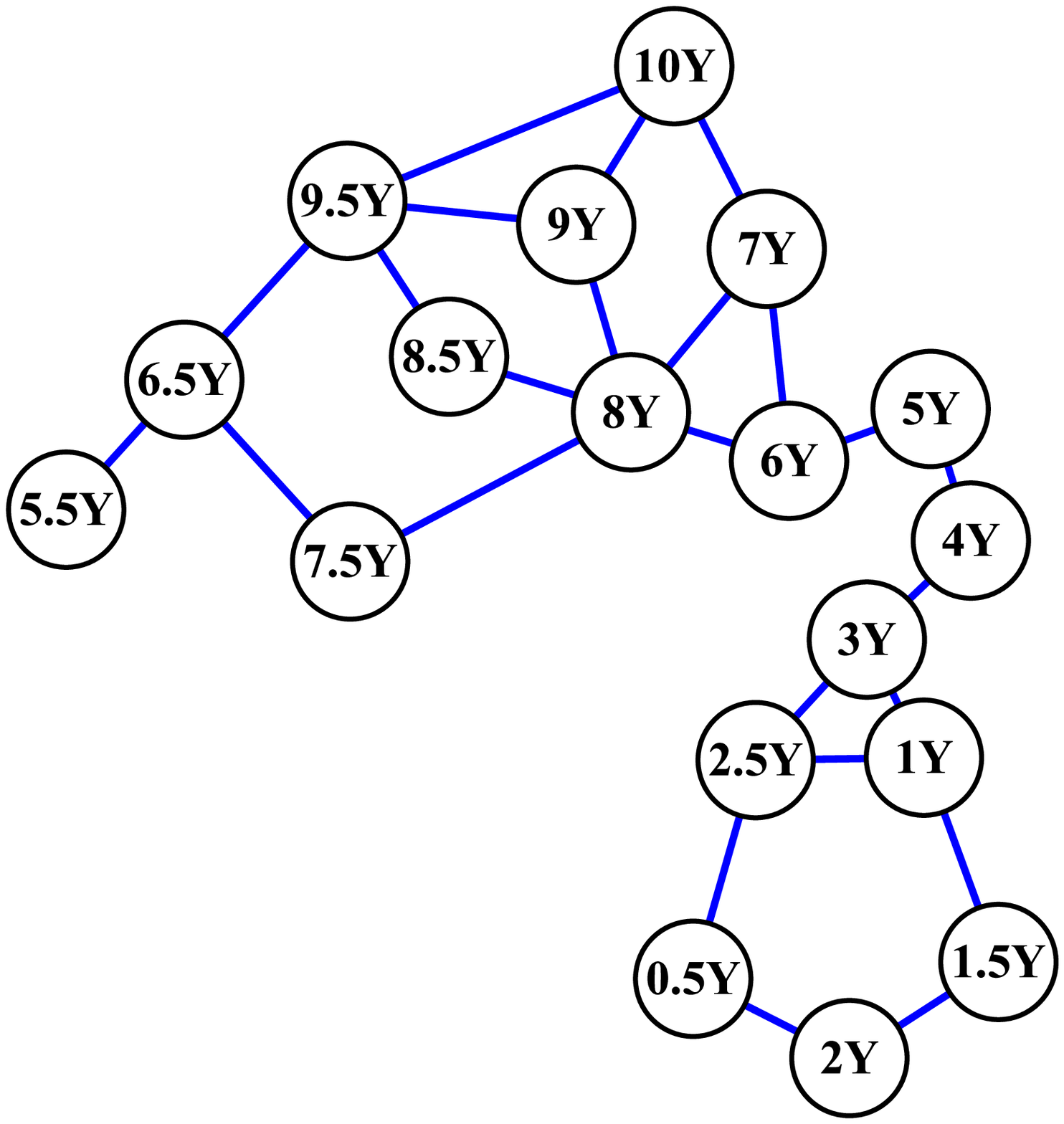}  \quad \quad\hfill 
\caption{\emph{Left:} a chordal graphical model: no cycles of length greater than three. \emph{Right:} a non-chordal graphical model. \label{fig:chordal}}
\end{center}
\end{figure}

\paragraph{Exogenous dependence models.} 
In the general case where the noise terms are correlated, with $Z_t\sim\mathcal{N}(0,\Sigma)$ for a certain noise covariance $\Sigma$, and the dependence structure is partly exogenous, we need to estimate the parameter matrix $A$ directly from the data. In Section \ref{s:decomp}, we estimated the matrix $A$ in the vector autoregressive model (\ref{eq:ar1}) by regressing $S_t$ on $S_{t-1}$:
\[
\hat A = \left(S_{t-1}^TS_{t-1}\right)^{-1} S_{t-1}^TS_{t}.
\]
Here too, we can modify this estimation procedure in order to get a sparse model matrix $A$. Our aim is again to both stabilize the estimation and highlight key dependence relationships between $S_t$ and $S_{t-1}$. We replace the simple least-squares estimate above by a penalized one. We get the columns of $A$ by solving:
\BEQ\label{eq:a-lasso}
a_i=\argmin_x \|S_{it}-S_{t-1}x\|^2+\gamma \|x\|_1
\EEQ
in the variable $x\in\reals^n$, where the parameter $\lambda>0$ controls sparsity. This is known as the LASSO (see \cite{Tibs96}) and produces sparse least squares estimates.

\subsection{Canonical decomposition with penalized estimation} 
\label{ss:pen-decomp} We showed that covariance selection highlights networks of dependence among assets, and that penalized regression could be used to estimate sparse model matrices $A$. We now show under which conditions these results can be combined to extract information on the support of the canonical portfolios produced by the decompositions in Section \ref{s:decomp} from the graph structure of the covariance matrix $\Gamma$ and of the model matrix $A$. Because both covariance selection and the lasso are substantially cheaper numerically than the sparse decomposition techniques in Section \ref{s:algos}, our goal here is to use these penalized estimation techniques as preprocessing tools to narrow down the range of assets over which we look for mean reversion. 

In Section \ref{s:decomp}, we saw that the \cite{Box77} decomposition for example, could be formed by solving the following generalized eigenvalue problem:
\[
\det(\lambda \Gamma - A^T \Gamma A) =0,
\]
where $\Gamma$ is the covariance matrix of the assets $S_t$ and $A$ is the model matrix in (\ref{eq:ar1}). Suppose now that 
our penalized estimates of the matrices $\Gamma$ and $A^T \Gamma A$ have disconnected graphs with identical clusters, i.e. have the same block diagonal structure, \cite[Th. 6.1]{Gilb94} shows that the support of the generalized eigenvectors of the pair $\{\Gamma,A^T \Gamma A\}$ must be fully included in one of the clusters of the graph of the inverse covariance $\Gamma^{-1}$. In other words, if the graph of the penalized estimate of $\Gamma^{-1}$ and $A$ are disconnected along the same clusters, then optimally unpredictable (or predictable) portfolios must be formed exclusively of assets in a single cluster.

This suggests a simple procedure for finding small mean reverting portfolios in very large data sets. We first estimate a sparse inverse covariance matrix by solving the covariance selection problem in (\ref{eq:sparseml-relax}), setting $\rho$ large enough so that the graph of $\Gamma^{-1}$ is split into sufficiently small clusters. We then check if either the graph is chordal or if penalized estimates of $A$ share some clusters with the graph of $\Gamma^{-1}$. After this preprocessing step, we use the algorithms of Section~\ref{s:algos} to search these (much smaller) clusters of variables for optimal mean reverting (or momentum) portfolios.

\begin{figure}[tp]
\begin{center}
\psfrag{card}[t][b]{Cardinality}
\psfrag{sharpe}[b][t]{Sharpe Ratio}
\includegraphics[width=.50\textwidth]{./figures/RatesNetNewNotChordal.eps} \quad
\includegraphics[width=.40\textwidth]{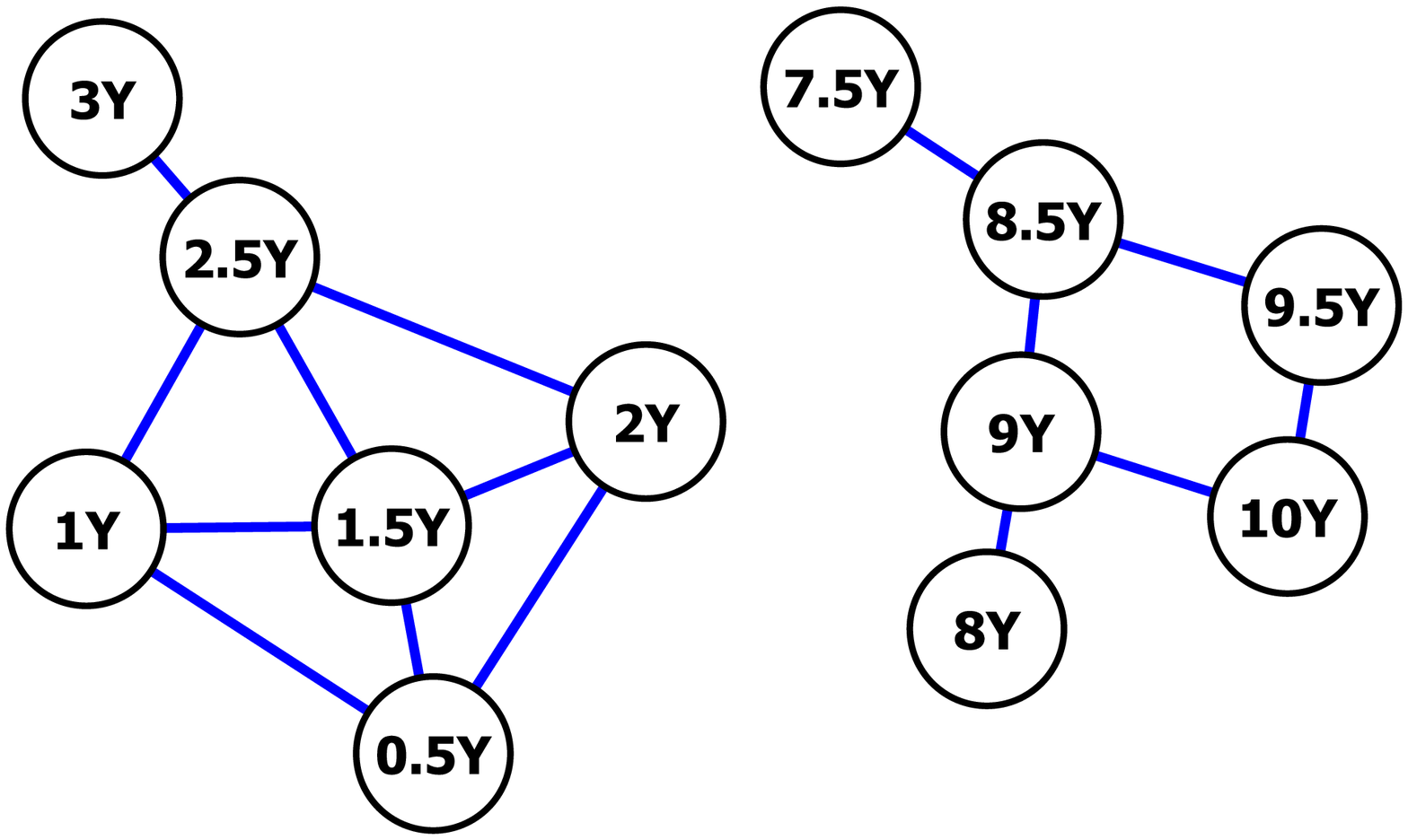}  \quad \quad\hfill 
\caption{\emph{Left:} a connected graphical model. \emph{Right:} disconnected models. \label{fig:chordal}}
\end{center}
\end{figure}

\section{Empirical results}
\label{s:res} In this section, we first compare the performance of the algorithms described in Section \ref{s:algos}. We then study the mean reversion versus sparsity tradeoff on various financial instruments. Finally, we test the performance of convergence trading strategies on sparse mean reverting portfolios.

\subsection{Numerical performance}

\begin{figure}[!h]
\begin{center}
\begin{tabular}{cc}
\psfrag{card}[t][b]{Cardinality}
\psfrag{meanrev}[b][t]{Mean Reversion}
\includegraphics[width=.48\textwidth]{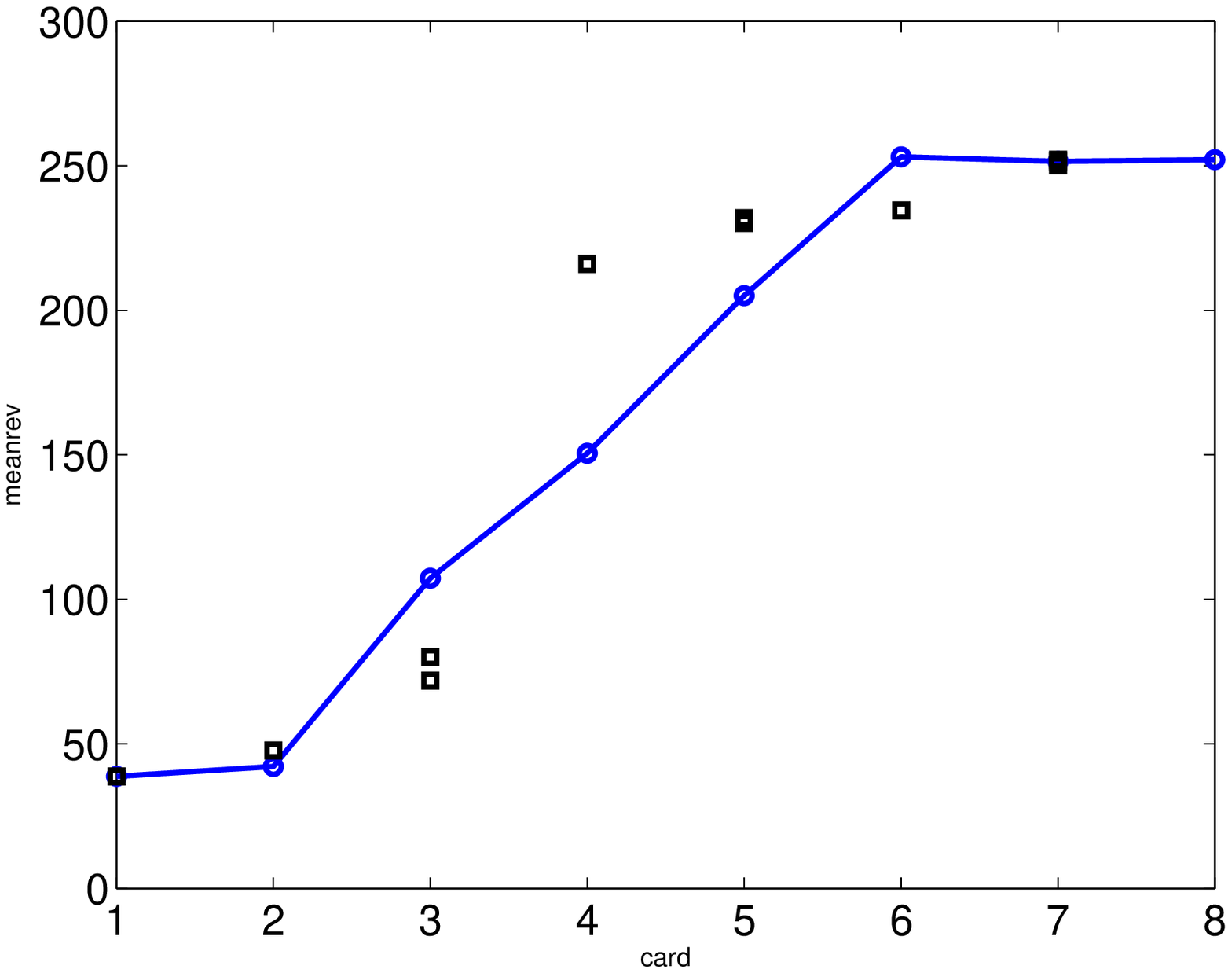} &
\psfrag{n}[t][b]{Total number of assets}
\psfrag{cpu}[b][t]{CPU time (in seconds)}
\includegraphics[width=.48\textwidth]{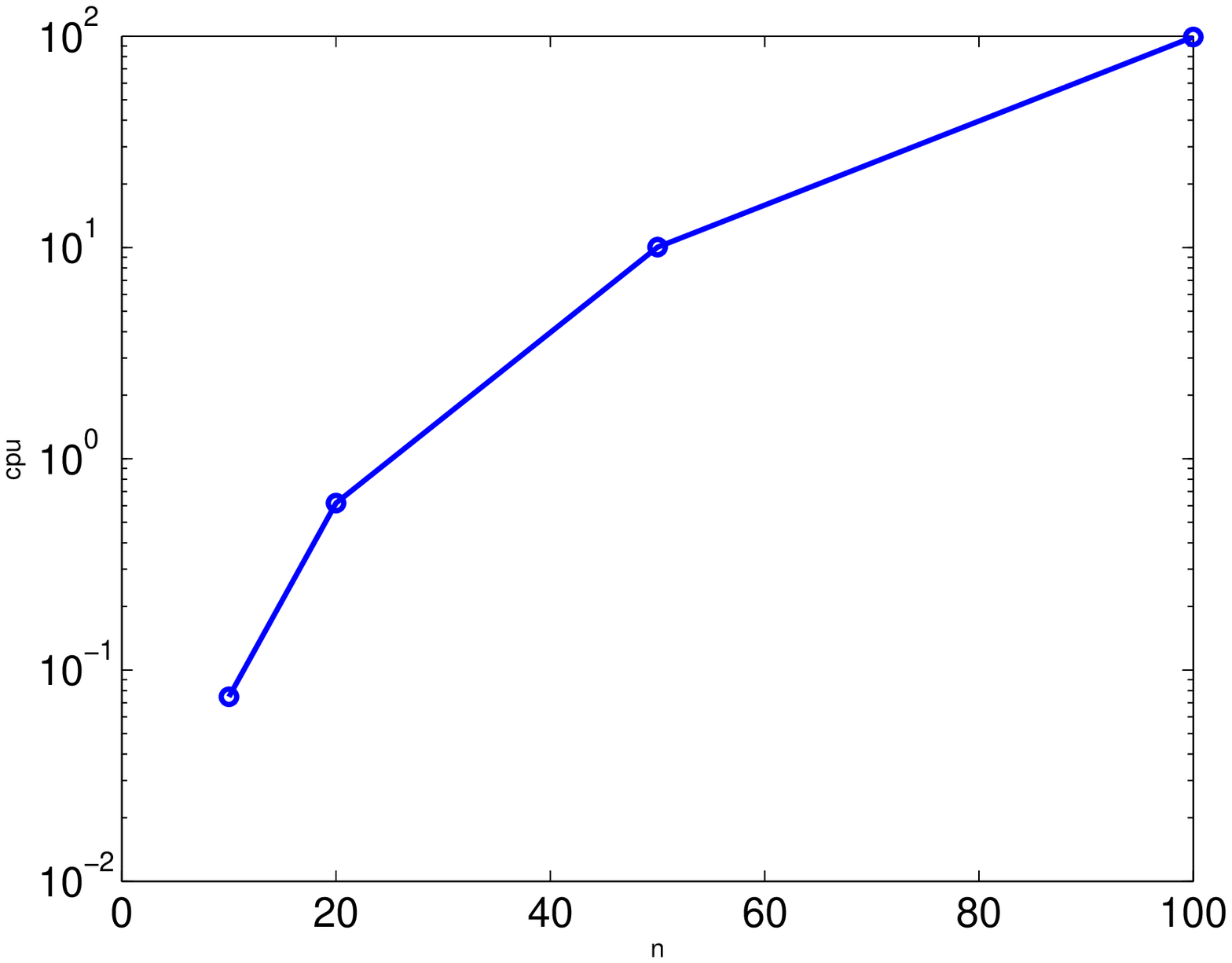}\\
\end{tabular}
\caption{\emph{Left:} Mean reversion coefficient $\lambda$ versus portfolio cardinality (number of nonzero coefficients) using the greedy search (circles, solid line) and the semidefinite relaxation (squares) algorithms on U.S. swap rate data. \emph{Right:} CPU time (in seconds) versus total number of assets $n$ to compute a full set of sparse portfolios (with cardinality ranging from 1 to $n$) using the greedy search algorithm.
\label{fig:gsdp}}
\end{center}
\end{figure}

\begin{figure}[p]
\begin{center}
\includegraphics[width=\textwidth]{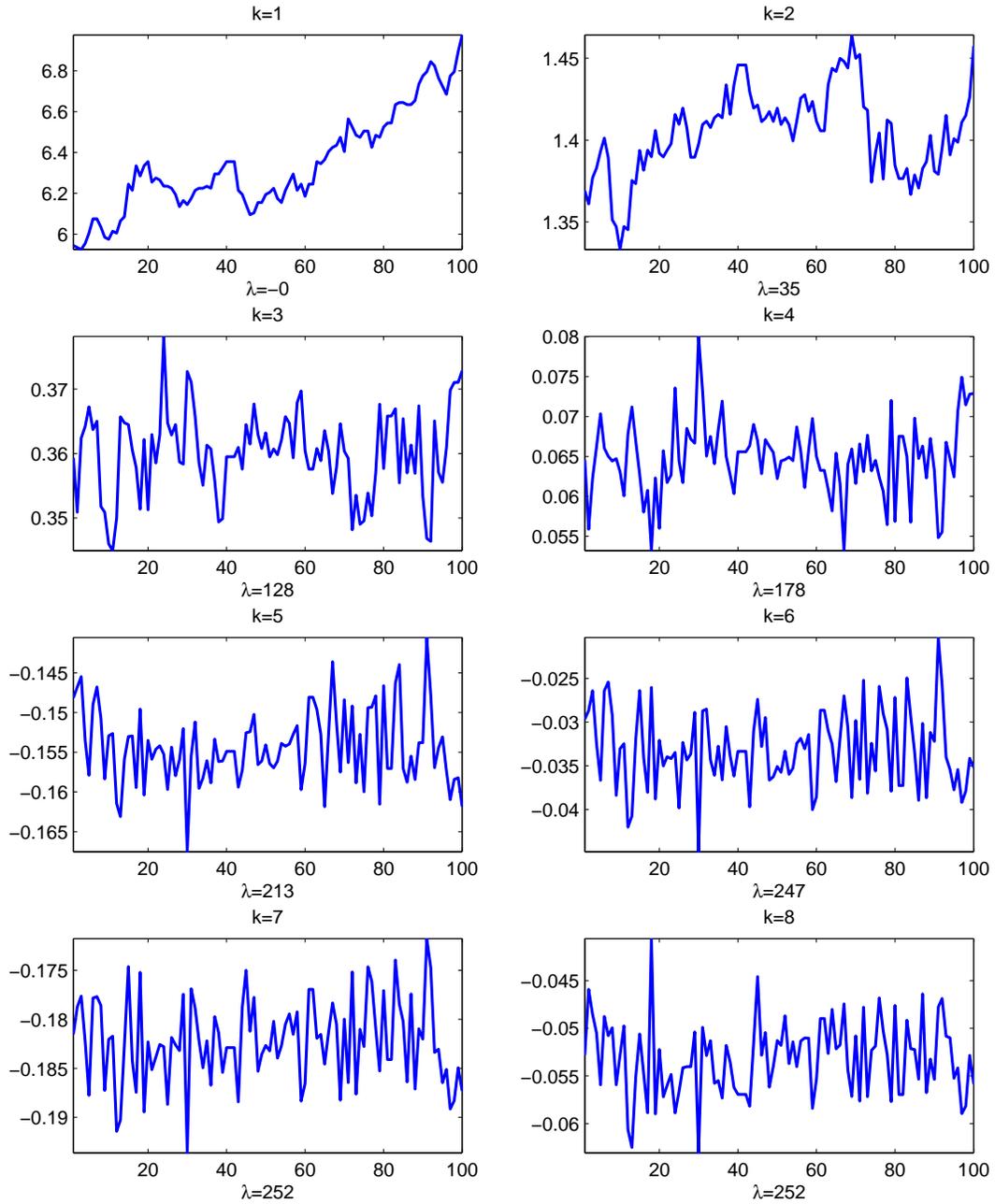} \caption{Sparse canonical decomposition on 100 days of U.S. swap rate data (in percent). The number of nonzero coefficients in each portfolio vector is listed as $k$ on top of each subplot, while the mean reversion coefficient $\lambda$ is listed below each one. \label{fig:sbt}}
\end{center}
\end{figure}

In Figure~\ref{fig:bt} we plotted the result of the Box-Tiao decomposition on U.S. swap rate data (see details below). Each portfolio is a \emph{dense} linear combination of swap rates, ranked in decreasing order of predictability. In Figure~\ref{fig:sbt}, we apply the greedy search algorithm detailed in Section \ref{s:algos} to the same data set and plot the \emph{sparse} portfolio processes for each target number of assets. Each subplot of Figure~\ref{fig:sbt} lists the number $k$ of nonzero coefficients of the corresponding portfolio and its mean reversion coefficient $\lambda$. 
Figure~\ref{fig:gsdp} then compares the performance of the greedy search algorithm versus the semidefinite relaxation derived in Section~\ref{s:algos}. On the left, for each algorithm, we plot the mean reversion coefficient $\lambda$ versus portfolio cardinality (number of nonzero coefficients). We observe on this example that while the semidefinite relaxation does produce better results in some instances, the greedy search is more reliable. Of course, both algorithms recover the same solutions when the target cardinality is set to $k=1$ or $k=n$. On the right, we plot CPU time (in seconds) as a function of the total number of assets to search. As a quick benchmark, producing 100 sparse mean reverting portfolios for each target cardinality between 1 and 100 took one minute and forty seconds.

\subsection{Mean reversion versus sparsity}
In this section, we study the mean reversion versus sparsity tradeoff on several data sets. We also test the persistence of this mean reversion out of sample.  

\begin{figure}[p]
\begin{center}
\psfrag{card}[t][b]{Cardinality}
\psfrag{meanrev}[b][t]{Mean Reversion}
\psfrag{bprange}[b][t]{Range (in bp)}
\includegraphics[width=.48\textwidth]{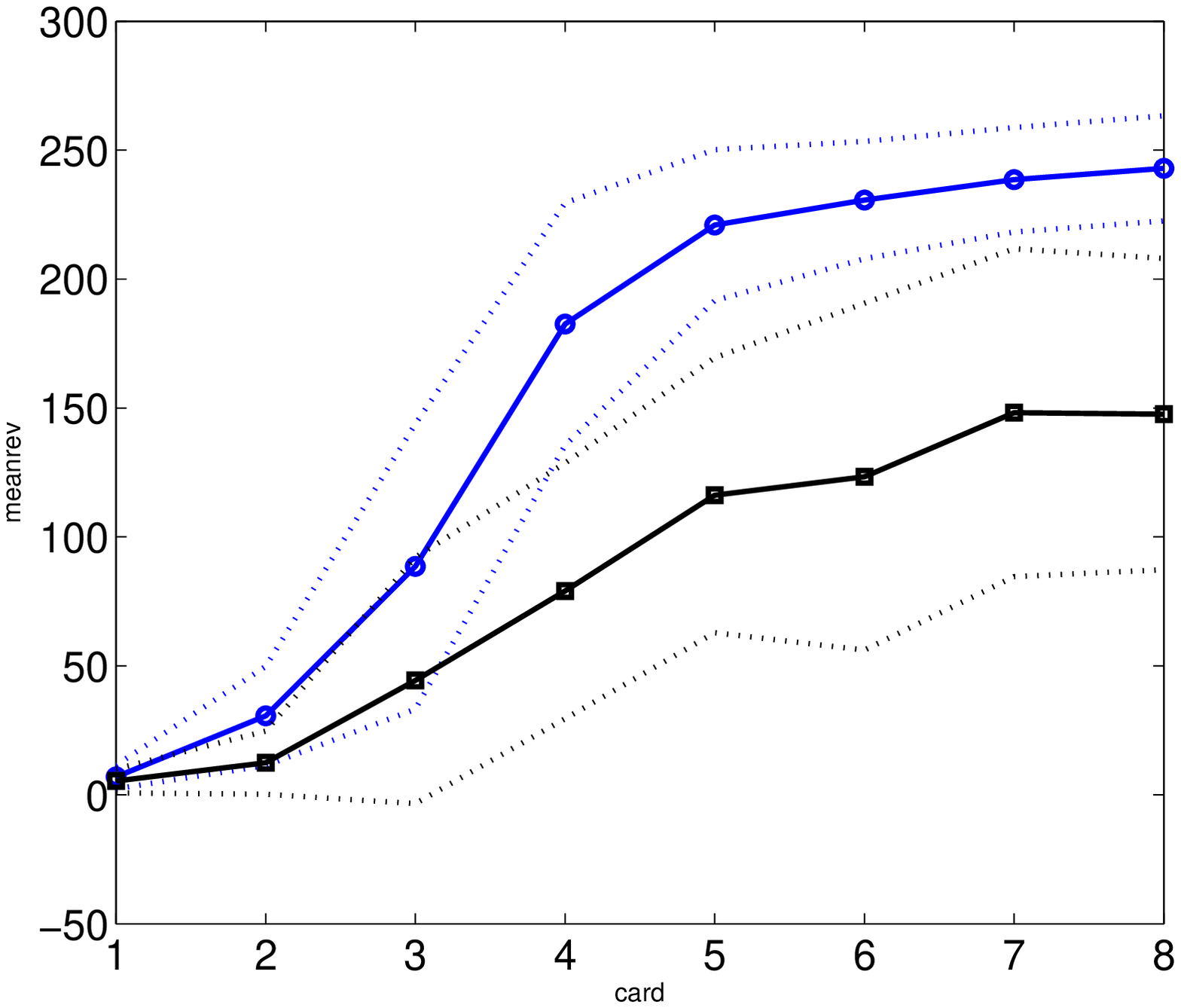} \hfill
\includegraphics[width=.48\textwidth]{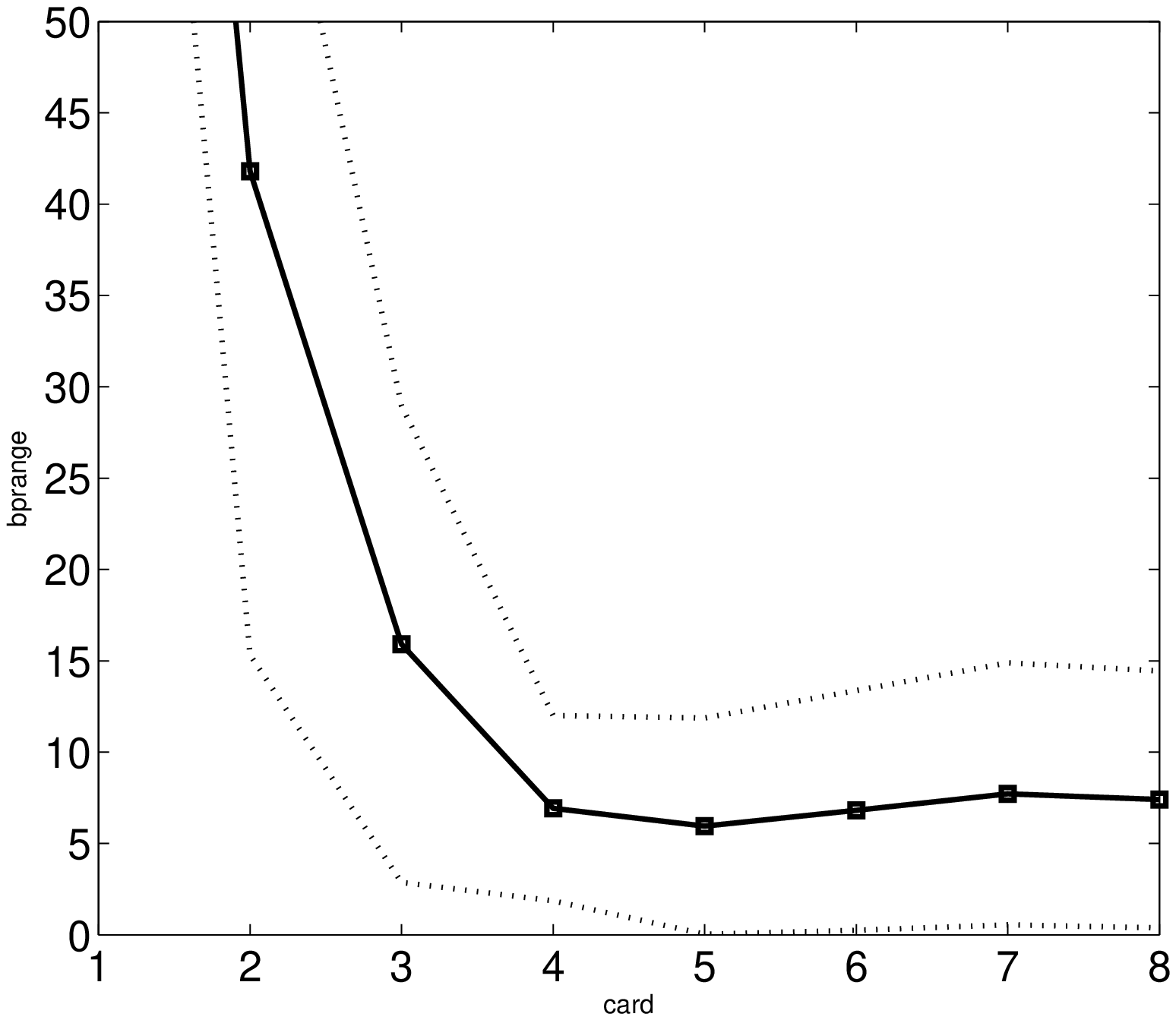}
\caption{\emph{Left:} mean reversion coefficient $\lambda$ versus portfolio cardinality (number of nonzero coefficients), in sample (blue circles) and out of sample (black squares) on U.S. swaps. \emph{Right:} out of sample portfolio price range (in basis points) versus cardinality (number of nonzero coefficients) on U.S. swap rate data. The dashed lines are at plus and minus one standard deviation. \label{fig:pers}}
\end{center}
\end{figure}

\paragraph{Swap rates.}
In Figure \ref{fig:pers} we compare in and out of sample estimates of the mean reversion versus cardinality tradeoff. We study U.S. swap rate data for maturities 1Y, 2Y, 3Y, 4Y, 5Y, 7Y, 10Y and 30Y from 1998 until 2005. We first use the greedy algorithm of Section~\ref{s:algos} to compute optimally mean reverting portfolios of increasing cardinality for time windows of 200 days and repeat the procedure every 50 days. We plot average mean reversion versus cardinality in Figure \ref{fig:pers} on the left. We then repeat the procedure, this time computing the (out of sample) mean reversion in the 200 days time window immediately following our sample and also plot average mean reversion versus cardinality. In Figure \ref{fig:pers} on the right, we plot the out of sample portfolio price range (spread between min. and max. in basis points) versus cardinality (number of nonzero coefficients) on the same U.S. swap rate data. Table \ref{tab:swap-port} shows the portfolio composition for each target cardinality.

\begin{table}[!h]
\small{
\begin{center}
\begin{tabular}{r|cccccccc}
&1&2&3&4&5&6&7&8\\ \hline
1Y & 0 & 0 & 0 & -0.041 & -0.037 & 0.036 & -0.013 & 0.001 \\
2Y & 0 & 0 & 0 & 0 & 0 & 0 & -0.102 & 0.117 \\
3Y & 0 & 0 & -0.288 & 0.433 & 0.419 & -0.437 & 0.547 & -0.495 \\
4Y & 0 & -0.714 & 0.806 & -0.803 & -0.802 & 0.809 & -0.767 & 0.702 \\
5Y & 1.000 & 0.700 & -0.517 & 0.408 & 0.424 & -0.389 & 0.317 & -0.427 \\
7Y & 0 & 0 & 0 & 0 & 0 & 0 & 0 & 0.219 \\
10Y & 0 & 0 & 0 & 0 & 0 & -0.031 & 0.025 & -0.130 \\
30Y & 0 & 0 & 0 & 0 & -0.007 & 0.016 & -0.008 & 0.014 \\
\end{tabular}
\caption{Composition of optimal swap portfolios for various target cardinalities.\label{tab:swap-port}}
\end{center}}
\end{table}

\paragraph{Foreign exchange rates.}
We study the following U.S. dollar exchange rates: Argentina, Australia, Brazil, Canada, Chile, China, Colombia, Czech Republic, Egypt, Eurozone, Finland, Hong Kong, Hungary, India, Indonesia, Israel, Japan, Jordan, Kuwait, Latvia, Lithuania, Malaysia, Mexico, Morocco, New Zealand, Norway, Pakistan, Papua NG, Peru, Philippines, Poland, Romania, Russia, Saudi Arabia, Singapore, South Africa, South Korea, Sri Lanka, Switzerland, Taiwan, Thailand, Turkey, United Kingdom, Venezuela, from April 2002 until April 2007. Note that exchange rates are quoted with four digits of accuracy (pip size), with bid-ask spreads around $0.0005$ for key rates.

\begin{figure}[!ht]
\begin{center}
\includegraphics[width=.6\textwidth]{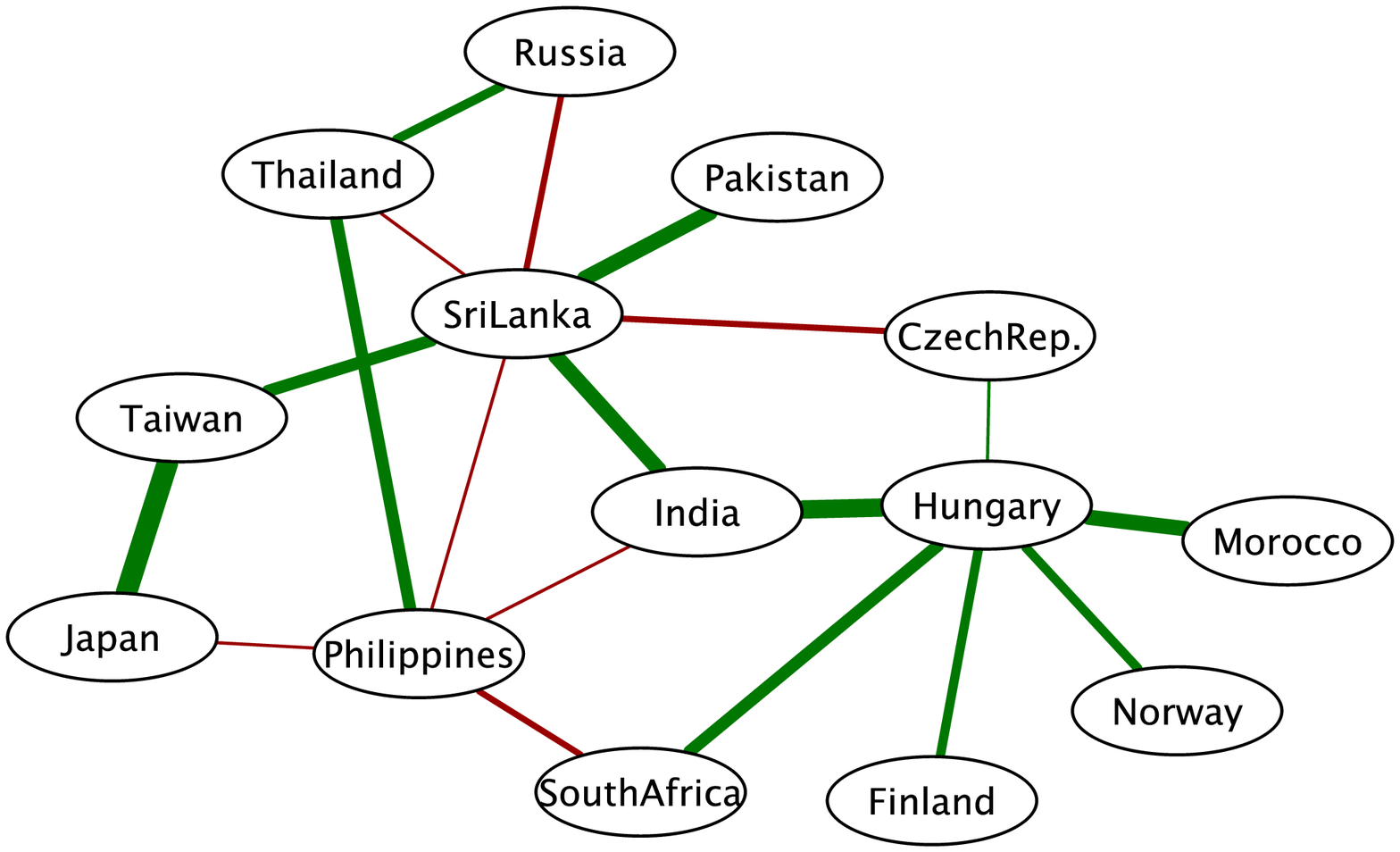} \caption{Graph of conditional covariance among a cluster of U.S. dollar exchange rates. Positive  dependencies are plotted as green links, negative ones in red, while the thickness reflects the magnitude of the covariance. \label{fig:forex-graph}}
\end{center}
\end{figure}

\begin{figure}[!ht]
\begin{center}
\psfrag{card}[t][b]{Cardinality}
\psfrag{meanrev}[b][t]{Mean Reversion}
\psfrag{bprange}[b][t]{Range}
\includegraphics[width=.48\textwidth]{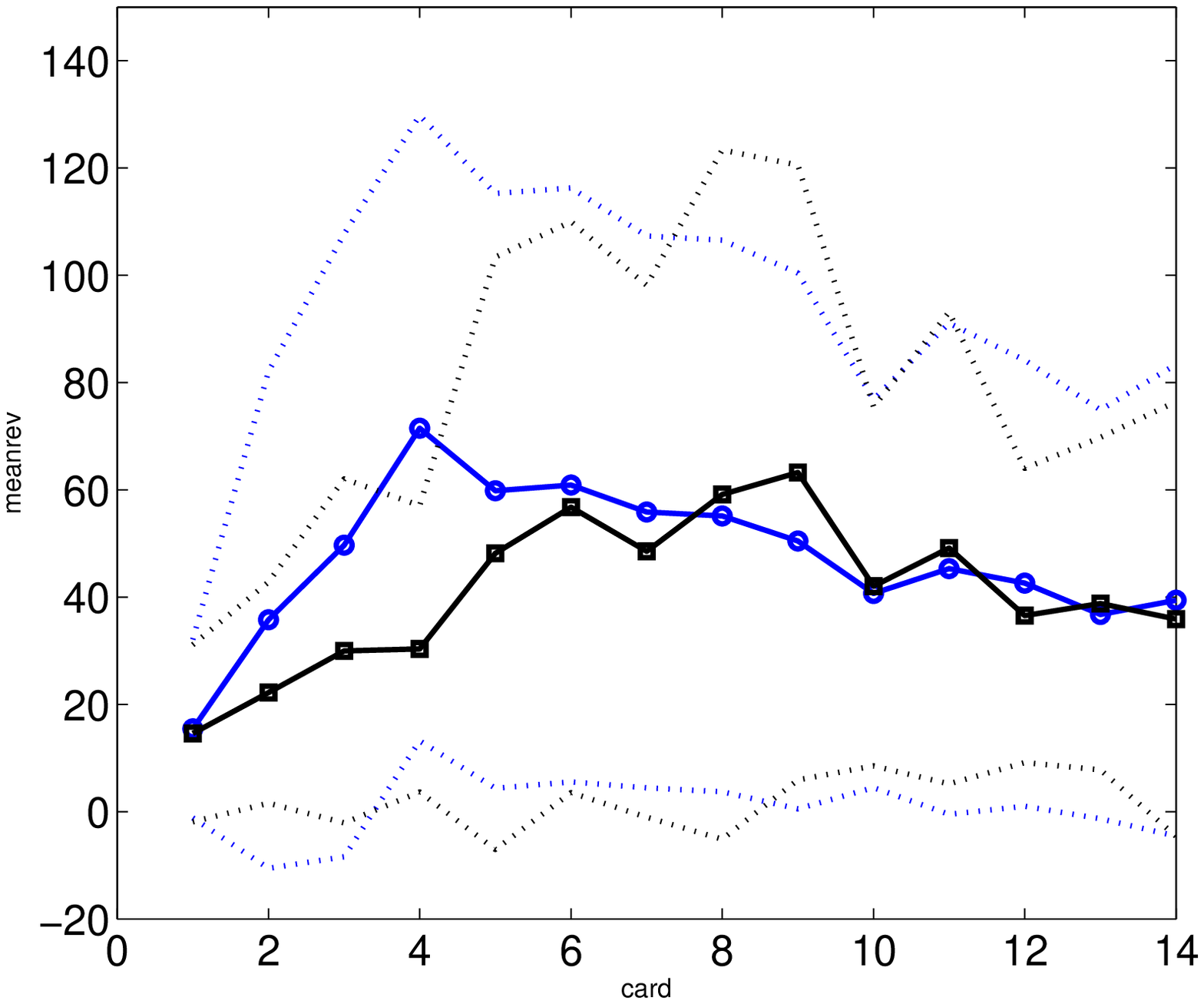} \hfill
\includegraphics[width=.48\textwidth]{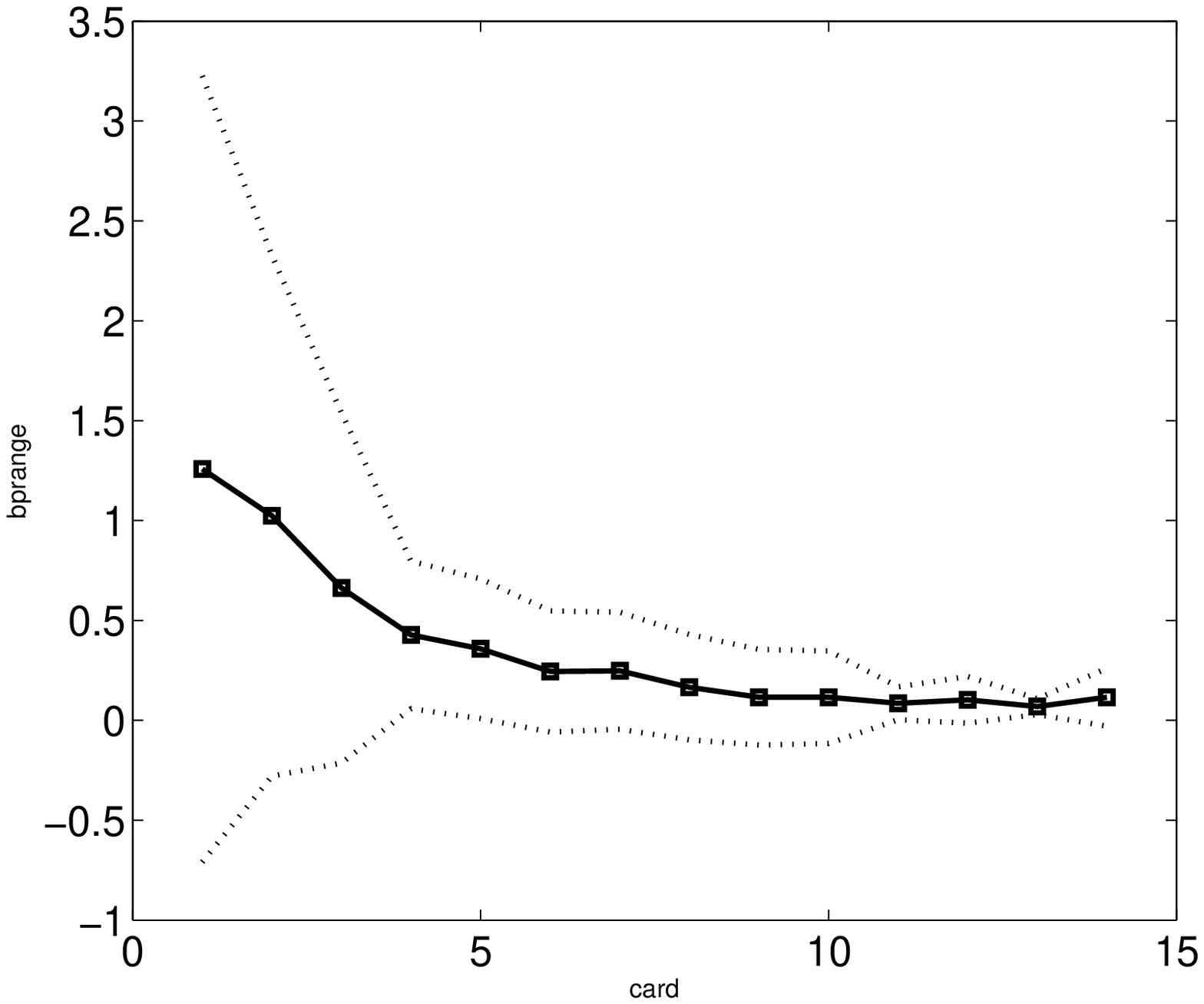}
\caption{\emph{Left:} out of sample mean reversion coefficient versus portfolio cardinality (number of nonzero coefficients), on 14 U.S. dollar exchange rates clustered by covariance selection. The sparse canonical decomposition was performed on both unpenalized estimates (black squares) and penalized ones (blue circles). \emph{Right:} out of sample portfolio price range (in percent) versus cardinality. The dashed lines are at plus and minus one standard deviation. \label{fig:OutForex}}
\end{center}
\end{figure}

After forming the sample covariance matrix $\Sigma$ of these rates, we solve the covariance selection problem in (\ref{eq:sparseml-relax}). This penalized maximum likelihood estimation problem isolates a cluster of 14 rates and we plot the corresponding graph of conditional covariances in Figure \ref{fig:forex-graph}. For these 14 rates, we then study the impact of penalized estimation of the matrices $\Gamma$ and $A$ on out of sample mean reversion. In Figure \ref{fig:OutForex}, we plot out of sample mean reversion coefficient $\lambda$ versus portfolio cardinality, on 14 rates selected by covariance selection. The sparse canonical decomposition was performed on both unpenalized estimates and penalized ones. The covariance matrix was estimated by solving the covariance selection problem (\ref{eq:sparseml-relax}) with $\rho=0.01$ and the matrix $A$ in (\ref{eq:ar1}) was estimated by solving problem (\ref{eq:a-lasso}) with the penalty $\gamma$ set to zero out $20\%$ of the regression coefficients.

We notice in Figure \ref{fig:OutForex} that penalization has a double impact. First, the fact that sparse portfolios have a higher out of sample mean reversion than dense ones means that penalizing for sparsity helps prediction. Second, penalized estimates of $\Gamma$ and $A$ also produce higher out of sample mean reversion than unpenalized ones. In Figure \ref{fig:OutForex} on the right, we plot portfolio price range versus cardinality and notice that here too sparse portfolios have a significantly broader range of variation than dense ones.

\subsection{Convergence trading}
Here, we measure the performance of the convergence trading strategies detailed in the appendix. In Figure \ref{fig:sharpeswap} we plot average out of sample sharpe ratio versus portfolio cardinality on a 50 days (out of sample) time window immediately following the 100 days over which we estimate the process parameters. Somewhat predictably in the very liquid U.S. swap markets, we notice that while out of sample Sharpe ratios look very promising in frictionless markets, even minuscule transaction costs (a bid-ask spread of 1bp) are sufficient to completely neutralize these market inefficiencies.

\begin{figure}[p]
\begin{center}
\psfrag{card}[t][b]{Cardinality}
\psfrag{sharpe}[b][t]{Sharpe Ratio}
\includegraphics[width=.49\textwidth]{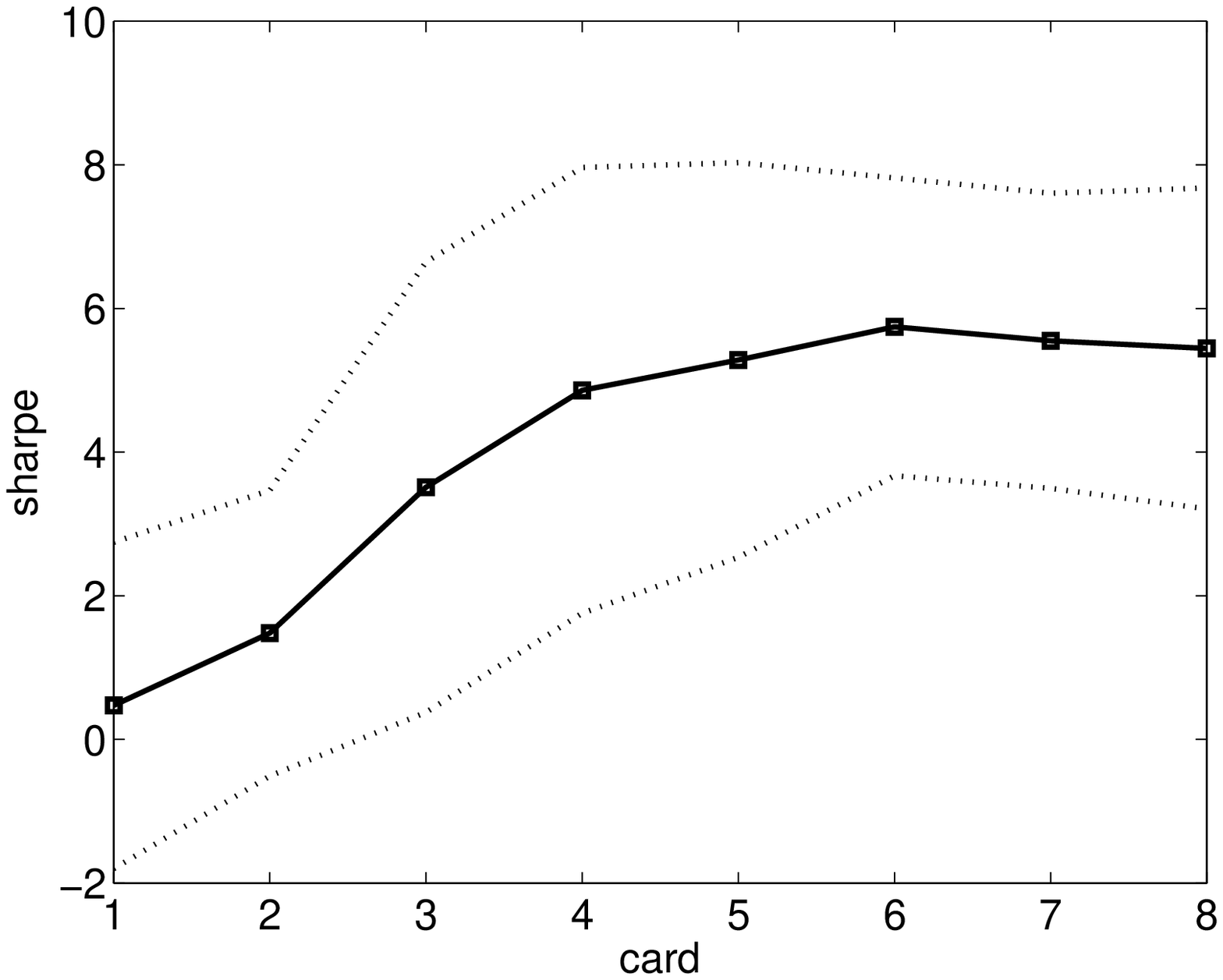} \hfill 
\includegraphics[width=.49\textwidth]{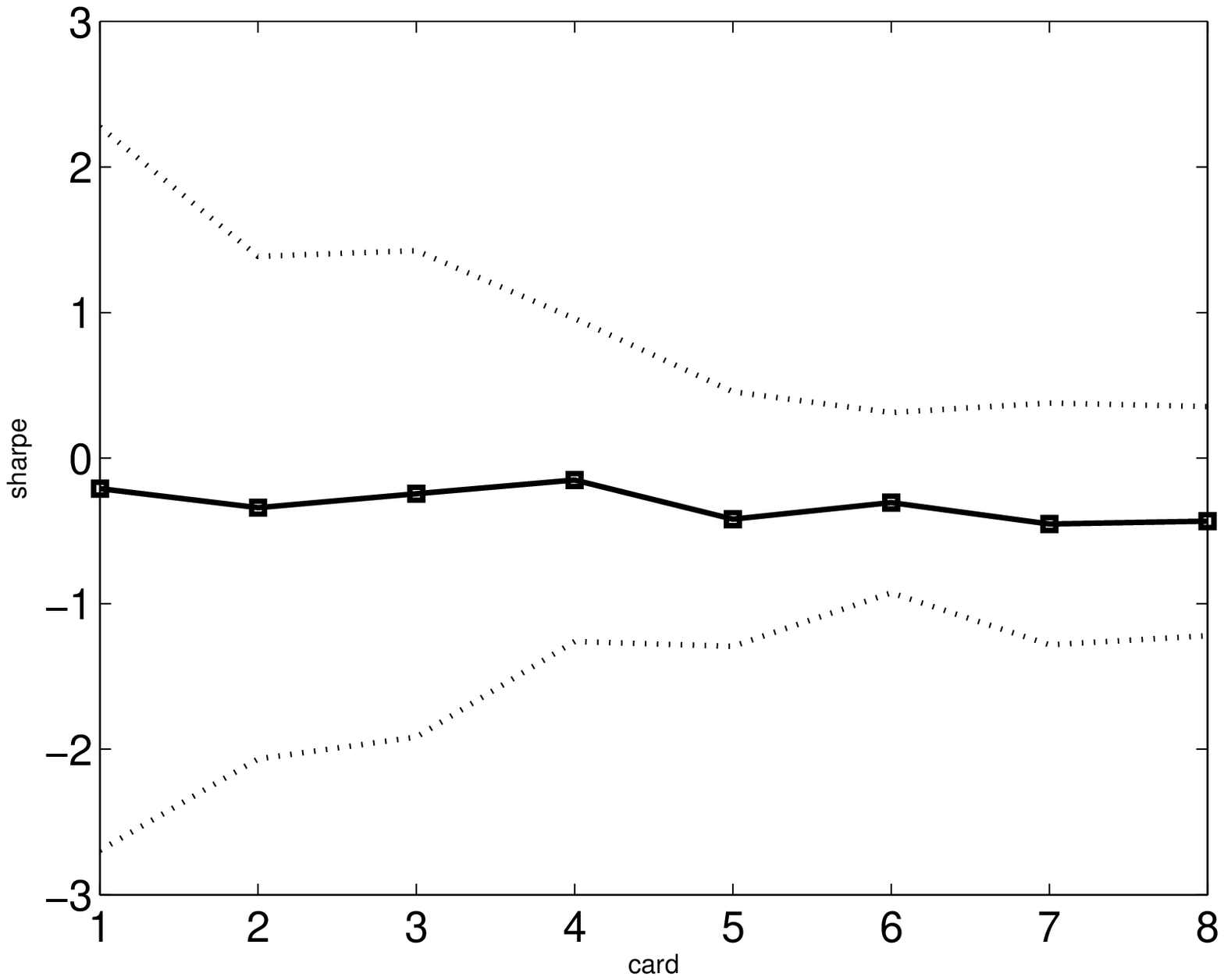} \caption{\emph{Left:} average out of sample sharpe ratio versus portfolio cardinality on U.S. swaps. \emph{Right:} idem, with a bid-ask spread of 1bp. The dashed lines are at plus and minus one standard deviation. \label{fig:sharpeswap}}
\end{center}
\end{figure}

\section{Conclusion}
We have derived two simple algorithms for extracting sparse (i.e. small) mean reverting portfolios from multivariate time series by solving a penalized version of the canonical decomposition technique in \cite{Box77}. Empirical results suggest that these small portfolios present a double advantage over their original dense counterparts: sparsity means lower transaction costs and better interpretability, it also improved out-of-sample predictability in the markets studied in Section \ref{s:res}. Several important issues remain open at this point. First, it would be important to show consistency of the variable selection procedure: assuming we know a priori that only a few variables have economic significance (i.e. should appear in the optimal portfolio), can we prove that the sparse canonical decomposition will recover them? Very recent consistency results by \cite{Amin08} on the sparse principal component analysis relaxation in \cite{dasp04a} seem to suggest that this is likely, at least for simple models. Second, while the dual of the semidefinite relaxation in (\ref{eq:max-meanrev-relax}) provides a bound on suboptimality, we currently have no procedure for deriving simple bounds of this type for the greedy algorithm in Section \ref{ss:greedy}. 

\section*{Acknowledgements}
The author would like to thank Marco Cuturi, Guillaume Boulanger and conference participants at the third Cambridge-Princeton conference and the INFORMS 2007 conference in Seattle for helpful comments and discussions. The author would also like to acknowledge support from NSF grant DMS-0625352, ONR grant number N00014-07-1-0150, a Peek junior faculty fellowship and a gift from Google, Inc.

\section*{Appendix}
In the previous sections, we showed how to extract small mean reverting (or momentum) portfolios from multivariate asset time series. In this section we assume that we have identified such a mean reverting portfolio and model its dynamics given by:
\BEQ\label{eq:ou}
dP_t= \lambda(\bar P- P_t) dt + \sigma dZ_t,
\EEQ
In this section, we detail how to optimally trade these portfolios under various assumptions regarding market friction and risk-management constraints. We begin by quickly recalling results on estimating the Ornstein-Uhlenbeck dynamics in (\ref{eq:ou}).

\subsection*{Estimating Ornstein-Uhlenbeck processes}
By explicitly integrating the process $P_t$ in (\ref{eq:ou}) over a time increment $\Delta t$ we get:
\BEQ\label{eq:int-ou}
P_{t}=\bar P + e^{-\lambda \Delta t}(P_{t-\Delta t}-\bar P)+\sigma \int_{t-\Delta t}^{t} e^{\lambda (s-t)}dZ_s,
\EEQ
which means that we can estimate $\lambda$ and $\sigma$ by simply regressing $P_t$ on $P_{t-1}$ and a constant. With 
\[
\int_{t-\Delta t}^{t} e^{\lambda (s-t)}dZ_s \sim \sqrt{\frac{1-e^{-2 \lambda \Delta t}}{2\lambda}}~\mathcal{N}(0,1),\] 
we get the following estimators for the parameters of $P_t$:
\BEAS
\hat\mu&=&\frac{1}{N}\sum_{i=0}^NP_{t}\\
\hat\lambda&=&-\frac{1}{\Delta t}\log\left(\frac{\sum_{i=1}^{N}(P_{t}-\hat\mu)(P_{t-1}-\hat\mu)}{\sum_{i=1}^{N}(P_{t}-\hat\mu)(P_{t}-\hat\mu)}\right)\\
\hat\sigma&=&\sqrt{\frac{2\lambda}{(1-e^{-2 \lambda \Delta t})(N-2)}\sum_{i=1}^N \left((P_{t}-\hat\mu)-e^{- \lambda \Delta t}(P_{t}-\hat\mu)\right)^2}\\
\EEAS
where $\Delta t$ is the time interval between times $t$ and $t-1$. The expression in (\ref{eq:int-ou}) also allows us to compute the \emph{half-life} of a market shock on $P_t$ as:
\BEQ\label{eq:hl}
\tau=\frac{\log 2}{\lambda},
\EEQ
which is a more intuitive measure of the magnitude of the portfolio's mean reversion.


\subsection*{Utility maximization in frictionless markets}
\label{ss:logu} Suppose now that an agent invests in an asset $P_t$ and in a riskless bond $B_t$ following:
\[
dB_t=rB_t dt,
\] 
the wealth $W_t$ of this agent will follow:
\[
dW_t=N_t dP_t + (W_t-N_tP_t)rdt.
\]
If $P_t$  follows a mean reverting process given by (\ref{eq:ou}), this is also:
\[
dW_t=(r(W_t-N_tP_t)+\lambda(\bar P- P_t)N_t)dt+N_t\sigma dZ_t.
\]
If we write the value function:
\[
V(W_t,P_t,t)=\max_{N_t}\textstyle\Expect_t\left[e^{-\beta(T-t)}U(W_t)\right],
\]
the H.J.B. equation for this problem can be written:
\BEAS
\beta V&=&\max_{N_t} \frac{\partial V}{\partial P} \lambda(\bar P_t -P_t)+\frac{\partial V}{\partial W}(r(W_t-N_tP_t)+\lambda(\bar P- P_t)N_t)+\frac{\partial V}{\partial t}\\
&&+\frac{1}{2}\frac{\partial^2 V}{\partial P^2}\sigma^2+\frac{1}{2}\frac{\partial^2 V}{\partial P\partial W}N_t\sigma^2+\frac{1}{2}\frac{\partial^2 V}{\partial W^2}N_t^2 \sigma^2
\EEAS
Maximizing in $N_t$ yields the following expression for the number of shares in the optimal portfolio:
\BEQ\label{eq:ns}
N_t=\frac{{\partial V}/{\partial W}}{{\partial^2 V}/{\partial W^2}\sigma^2}(\lambda(\bar P -P_t)-rP_t)-\frac{{\partial^2 V}/{\partial P\partial W}}{{\partial^2 V}/{\partial W^2}}
\EEQ
\cite{Jure06} solve this equation explicitly for $U(x)=\log x$ and $U(x)=x^{1-\gamma}/(1-\gamma)$ and we recover in particular the classic expression:
\[
N_t=\left(\frac{\lambda(\bar P-P_t)-rP_t}{\sigma^2}\right) W_t,
\]
in the log-utility case.

\subsection*{Leverage constraints}
\label{ss:leverage} Suppose now that the portfolio is subject to fund withdrawals so that the total wealth evolves according to:
\[
dW=d\Pi+dF
\]
where $d\Pi=N_t dP_t + (W_t-N_tP_t)rdt$ and $dF$ represents fund flows, with:
\[
dF=fd\Pi+\sigma_fdZ_t^{(2)}
\]
where $Z_t^{(2)}$ is a Brownian motion (independent of $Z_t$). \cite{Jure06} show that the optimal portfolio allocation can also be computed explicitly in the presence of fund flows, with:
\[
N_t=\left(\frac{\lambda(\bar P-P_t)-rP_t}{\sigma^2}\right) \frac{1}{(1+f)}W_t =L_tW_t,
\]
in the log-utility case. Note that the constant $f$ can also be interpreted in terms of leverage limits. In steady state, we have:
\[
P_t \sim \mathcal{N}\left(\bar P, \frac{\sigma^2}{2\lambda}\right)
\]
which means that the leverage $L_t$ itself is normally distributed. If we assume for simplicity that $\bar P=0$, given the fund flow parameter $f$, the leverage will remain below the level $M$ given by:
\BEQ\label{eq:lev}
M= \frac{\alpha(\lambda + r)}{(1+f)\sigma\sqrt{2\lambda}}
\EEQ
with confidence level $N(\alpha)$, where $N(x)$ is the Gaussian CDF. The bound on leverage $M$ can thus be seen as an alternate way of identifying or specifying the fund flow constant  $f$ in order to manage capital outflow risks.

\small{
\bibliographystyle{agsm}
\bibliography{MainPerso}
}

\end{document}